\documentclass[12pt,preprint]{aastex}
\usepackage{natbib}
\usepackage{graphicx}
\usepackage{wasysym}
\usepackage{txfonts}
\def\fH2{\mbox {f$_{{\rm H}_2}$}}

\def\EBV{E(B-V)}

\def\nH2{\mbox{${\rm n}(\HH$)}}
\def\enH2{\mbox{$n_{(\HH$)}}}

\def\pccc{~{\rm cm}^{-3}} 
 
\def\pcc {\mbox{${~{\rm cm}^{-2}}$}}
\def\absb {\mbox{$|b|$}}

\def\Tsub#1 {\mbox{${\rm T}_{\rm #1}$}}
\def\TK  {\Tsub K }
\def\TB  {\Tsub B }

\def\Tsp {\Tsub sp }

 \def\arcmin{\mbox{$^{\prime}$}}

\def\degr{\mbox{$^{\rm o}$}}
\def\p{\mbox{$^+$}}

\def\h13cop{\mbox{{H$^{13}$CO\p}}}

\def\c3h2{\mbox{C$_3$H$_2$}}

 \def\R0{R$_0$}
\def\G0{\mbox{G$_0$}}

\def\ddeg{{}^\circ\kern-.1em}

\def\kms{\mbox{km\,s$^{-1}$}}

\def\E#1 {$10^{#1}$}
\def\E#1 {E{#1}}
\def\P#1,{$\nH2\TK~=~#1\times~10^4\pccc$~K}
\def\ec#1,#2,#3,{#1\,(#2)\E{#3}}

\def\H3{\mbox{H$_3$}}
\def\Lya{\mbox{Ly-$\alpha$}}

\def\RH2{\mbox{R$_{\rm G}$}}
\def\g13{\mbox{g$_{13}$}}

\def\kHeH2{\mbox{$k_{ He-\HH}$}}

\def\tim#1,#2{\mbox{{$#1\times10^{#2}$}}}

\def\WHI{$\Xi_{{\rm H I}}$}

\newcommand{\emm}[1]{\ensuremath{#1}}   
\newcommand{\emr}[1]{\emm{\mathrm{#1}}} 

\newcommand{\HH}{\emr{H_2}}

\slugcomment{generated \today}

\shorttitle{Red Dawn for H I}
\shortauthors{Harvey Liszt}

\begin{document}


\title{ \EBV, N(H I) and N(\HH)}


\author{Harvey Liszt}
\affil{National Radio Astronomy Observatory \\
        520 Edgemont Road, Charlottesville, VA 22903-2475}

\email{hliszt@nrao.edu}





\begin{abstract}

We consider the structure of the N(H I) - \EBV\ relationship when H I is measured
in the 21 cm radio line and \EBV\ is defined by far-IR dust-derived measures.
We derive reddening-dependent corrections to N(H I) based on interferometric 
absorption measurements over the past 30 years that follow a single power-law 
relationship $\int \tau(H I) dv = 14.07 ~\kms\ $\EBV$^{1.074}$ at 
0.02 $\la$ \EBV\ $\la$ 3 mag.  Corrections to 21cm line-derived H I column 
densities are too small to have had any effect on the ratio 
N(H I)/\EBV\ $= 8.3 \times 10^{21}\pcc$ mag$^{-1}$ we derived at 
0.015 $\la$ \EBV\ $\la$ 0.075 mag  and \absb\ $\ge$ 20\degr; they are 
also too small to explain 
the break in the slope of the N(H I) - \EBV\ relation at \EBV\ $\ga$ 0.1 mag
that we demonstrated around the Galaxy at \absb $\ge 20$\degr.  The latter must 
therefore be attributed to the onset of \HH-formation and we show that models of 
\HH\ formation in a low density diffuse molecular gas can readily explain the 
inflected N(H I)- \EBV\ relationship.  Below \absb\ = 20\degr\ 
N(H I)/\EBV\ measured at 0.015 $\la$ \EBV\ $\la$ 0.075 mag increases steadily 
down to \absb\ = 8\degr\ where sightlines with small \EBV\ no longer occur.
 By contrast, the ratio N(H I)/\EBV\ measured over 
all \EBV\ declines to N(H I)/\EBV\  $= 5-6 \times 10^{21}\pcc$ mag$^{-1}$
at \absb\ $\la 30$\degr, perhaps providing an explanation of the
difference between our results and the gas/reddening ratios measured previously 
using stellar spectra.

\end{abstract}



\section{Introduction}

In an earlier paper \citep{Lis14} we considered the relationship between 
H I column density N(H I) derived from the large-scale 21cm  H I sky 
surveys \citep{KalBur+05,PeeHei+11}, and reddening \EBV\ as derived from far 
IR dust emission by \cite{SchFin+98} (SFD98).   We traced N(H I) and \EBV\ 
around the sky
at galactic latitudes $\absb = 20\degr - 60\degr$ and considered data at lower
column densities $0.015 \la \EBV \la 0.075$ mag, where the hydrogen should 
be in the form of neutral atoms and corrections to N(H I) for
saturation and \HH-formation are unimportant.  We showed that the relevant 
value of the gas/reddening ratio is N(H I)/\EBV\ $= 8.3 \times 10^{21}\pcc$ 
mag$^{-1}$, considerably larger than the usually-cited ratio
N(H)/\EBV\ $= 5.8 \times 10^{21}\pcc$ mag$^{-1}$ derived from optical/uv
absorption measurements toward stars by \cite{BohSav+78}, where N(H) = N(H I) +
2N(\HH).  It is also larger than the values 
N(H I)/\EBV\ $= 4.8-5.2 \times 10^{21}\pcc$ mag$^{-1}$ that are consistently
quoted for stellar reddening and \Lya\ absorption toward early-type stars
\citep{BohSav+78,ShuVan85,DipSav94}.

\begin{figure*}
\includegraphics[height=6.2cm]{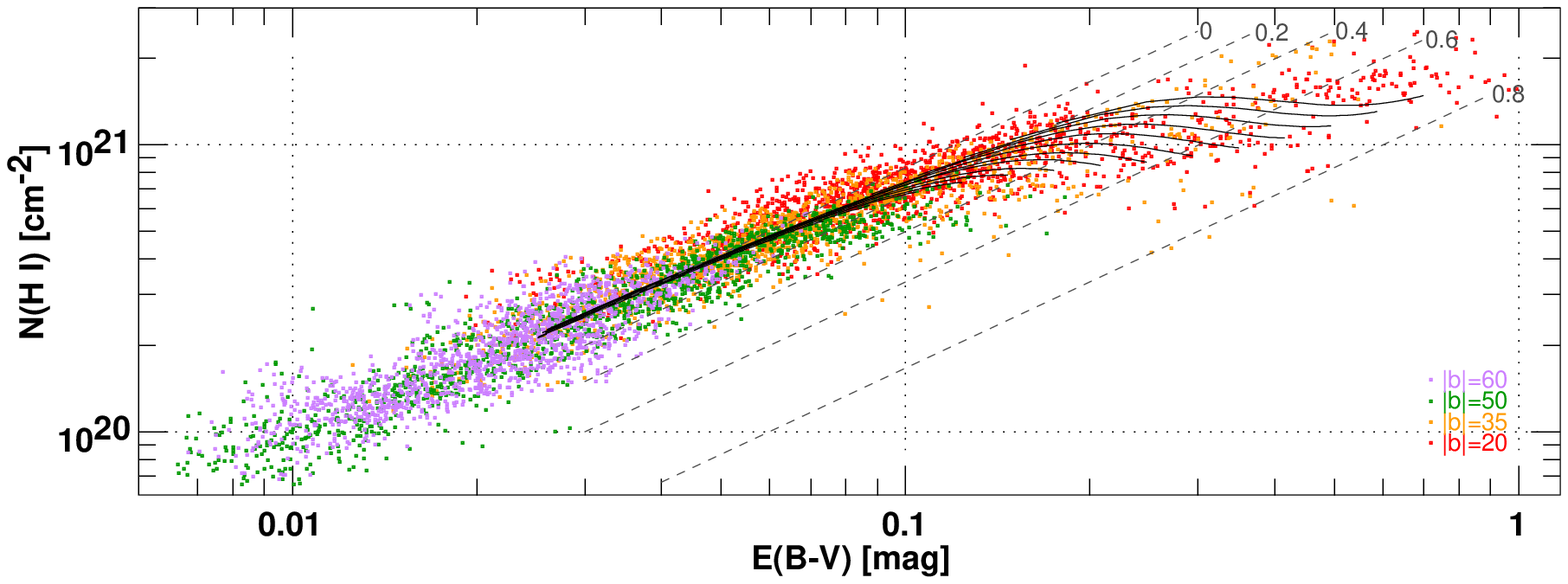}
 \caption[] {Reddening \EBV\ from SFD98 and opacity-corrected LAB-survey H I 
column density N(H I).  Data were sampled on grid points of the LAB survey
 at even 0.5\degr\ intervals around the sky at b = $\pm$20\degr, $\pm$35\degr, 
$\pm$50\degr\ and $\pm$60\degr.  Values of N(H I) were corrected as
outlined in Sect. 2 and illustrated in Fig. 4.  Dashed lines indicate loci 
of fixed molecular fraction \fH2\ = 1-N(H I)/N(H) for N(H)/\EBV\ 
$= 8.3 \times 10^{21}\pcc$ mag$^{-1}$. Solid lines correspond to models of 
\HH-formation in  low-density diffuse gas as discussed in Sect. 3.
}
\end{figure*}

In fact there are two mysteries in the disparity between radio/IR and optical/uv
derivations of N(H)/\EBV.  The first is the numerical discrepancy, which puts several
21 cm surveys and the work of SFD98 together on one side in opposition to \Lya\ 
measurements by several groups using IUE and Copernicus and stellar reddening on 
the other.  The second is the very nearly constant value for N(H I)/\EBV\ quoted 
for the stellar data, as opposed to the radio-IR relationship shown in Figure 1 
of \cite{Lis14} that had a very strong point of inflection to smaller N(H I)/\EBV\ 
at \EBV\ $\ga$ 0.08 mag.  In principle the inflection in the radio data could 
reflect either the influence of saturation of the 21cm line profiles or the 
expected onset of \HH\ formation as originally discovered in the uv absorption data 
\citep{SavDra+77}.  We noted a seemingly similar inflection in the IUE results 
of \cite{DipSav94},  whereby much higher values of N(H I)/\EBV\ were seen at 
\EBV $< 0.1$ mag (their Fig. 4a), but this was not taken into account in their 
final result.  It should also have been present in the earlier treatment of the 
IUE results by \cite{ShuVan85}, who derived very nearly the same numerical result as
in the later work.

This work is largely concerned with understanding the change in slope of the radio-IR
defined N(H I)-\EBV\ relationship, unravelling the possibly competing effects of 21cm
H I line saturation and \HH-formation.  In Section 2 we discuss independent measures 
of 21cm H I optical depth as a function of reddening, which can be employed to show 
that saturation corrections to 21cm measurements of N(H I) are small at least until 
\EBV $>$ 0.3 mag.  In that case, only \HH-formation can explain the observed inflection.  
In Sect. 3 we show that rather  conventional models of \HH\ formation in a low density
diffuse molecular gas can reproduce the observed N(H I)-\EBV\ relationship.
In Section 4 we extend our analysis to lower galactic latitude \absb\ = 9\degr- 20\degr\
and show that that there are progressively higher values of N(H I)/\EBV\ at all \EBV\
as \absb\ declines.  Section 5 is a brief summary and discussion.

\section{Optical depth and saturation correction in the 21cm line}

Figure 1 shows an updated and annotated version of the N(H I)-\EBV\ relationship 
first shown in \citep{Lis14}, again using H I profiles from the LAB 
(Leiden-Argentina-Bonn) all-sky H I survey 
\citep{KalBur+05} and \EBV\ from the work of SFD98.  Explaining the annotations and the 
rather modest saturation correction to N(H I) that has been applied to the data is
the subject of the current work.  In passing, please note that Figure 1 of \cite{Lis14}
inadvertently displayed the data for only that half the sky at 
0\degr\ $\leq l \leq 180$\degr, without affecting the numerical results.  
Note also that the phenomenon of gas without
dust, which manifests itself as high ratios of N(H I)/\EBV\ at small \EBV, is
not clearly present down to reddenings as small as 0.01 mag.  Also missing at
small values of \EBV\ is a downturn in N(H I) that could have signalled the
presence of in increasing fracion of warm ionized gas.

\subsection{Systematic variation of the 21cm optical depth with reddening}

To understand the possible effects of saturation we began by binning the data 
in reddening (averaging over all the data comprising Fig. 1) and forming mean 
H I emission profiles as a function of reddening.  Figure 2 shows some of
these for 0.025 mag $\leq \EBV \leq 0.33$ mag and it is clear that imputing
high optical depth to the profiles around \EBV\ = 0.1 mag would require 
spin temperatures below 30 K for which there is no support in such
diffuse gas.

\begin{figure}
\includegraphics[height=9cm]{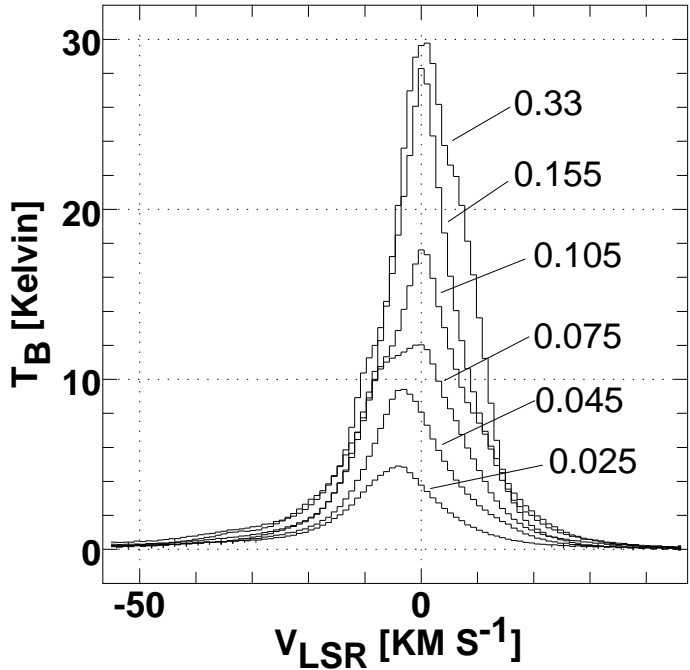}
 \caption[] {Mean H I profiles after binning the Fig. 1 data in \EBV\ 
around \EBV\ values \EBV\ = 0.025, 0.045, .. 0.33 mag as indicated.}
\end{figure}

The argument for modest optical depths  may be made quantitative by considering 
the variation of measured 21cm optical depth \WHI\ = $\int \tau(H I) dv$ 
(units of \kms) with reddening first discussed by \cite{LisPet+10} using a 
combination of their own more recent VLA data and that measured earlier by 
\cite{DicKul+83}.  \cite{LisPet+10} showed that there is a strong, nearly-linear relationship 
between \WHI\ and \EBV\ but with scant data at higher galactic latitude and
with much scatter and sparse data-coverage at \EBV $<$ 0.3 mag. This situation is 
alleviated by inclusion of the new results of \cite{RoyKan+13} as shown 
in Fig. 3 \footnote{ For sightlines in common between the two datasets, the value
from \cite{LisPet+10} has been retained}.

The error-weighted regression line in  Fig. 3 is log \WHI\ $= 1.147 \pm 0.019 
+ (1.074 \pm 0.034)$ log \EBV\ or \WHI\ = 14.07\EBV$^{1.074}$, using all the 
datapoints shown at \EBV $>$ 0.02 mag. The ratio \WHI/\EBV\ changes by only
45\% over the range 0.02 $\leq \EBV \leq 3$ mag.  

H I absorption (hence the presence of the 
cold neutral medium) is not consistently detected below \EBV\ = 0.02 mag.  
This was noted by \cite{KanBra+11} who described the lack of H I absorption 
for N(H I) $< 2\times 10^{20}\pcc$.  This implies a ratio 
N(H I)/\EBV\ $= 10^{22}\pcc$ mag$^{-1}$ in keeping with the values found in 
our work but a more quantitative estimate can be derived from the table of 
\WHI\ and optical-depth corrected N(H I) of \cite{RoyKan+13}, from which it 
is found that $<$N(H I)/\EBV$> = 7.7\pm 1.4 \times 10^{21}\pcc$ mag$^{-1}$.

Note that Fig. 3 appears to validate the use of \EBV\ from SFD98 up to rather 
higher values and at rather lower galactic latitudes than are usually believed 
to be reliable, see the discussion in the original work.  Unlike the original 
discussion in \cite{LisPet+10} the relationship between \WHI\ and \EBV\ is 
demonstrated over a wide range of galactic latitudes  1\degr-60\degr.

\subsection{Saturation correction to N(H I)}

\begin{figure}
\includegraphics[height=8.4cm]{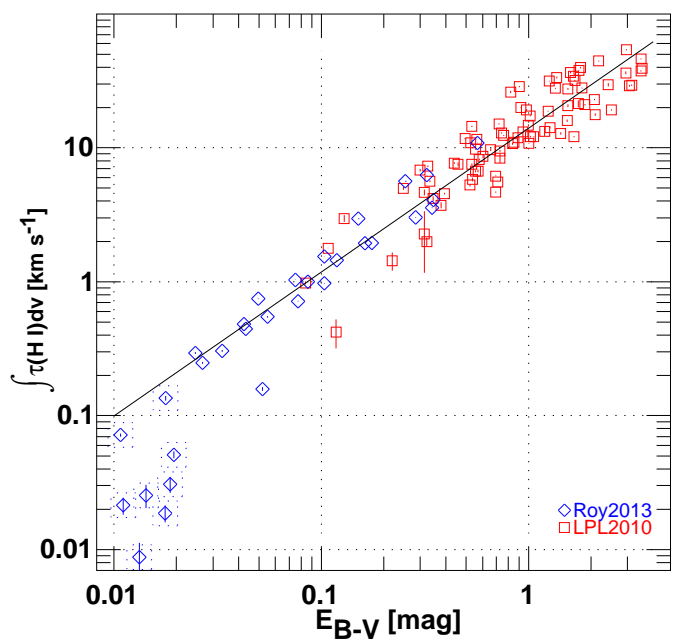}
 \caption[]{Velocity-integrated H I optical depth vs. reddening.  The H I
data are those of \cite{LisPet+10}, taken from their work and from 
\cite{DicKul+83}, and more recent results of \cite{RoyKan+13}.  The solid line 
is an error-weighted power-law fit to the data at \EBV\ $>$ 0.02 mag,
$\int \tau(H I) dv = 14.07 ~\kms\ $\EBV$^{1.074}$
Errors in \EBV, not shown, are 16\% according to SFD98.} 
\end{figure}

\begin{figure}
\includegraphics[height=9.8cm]{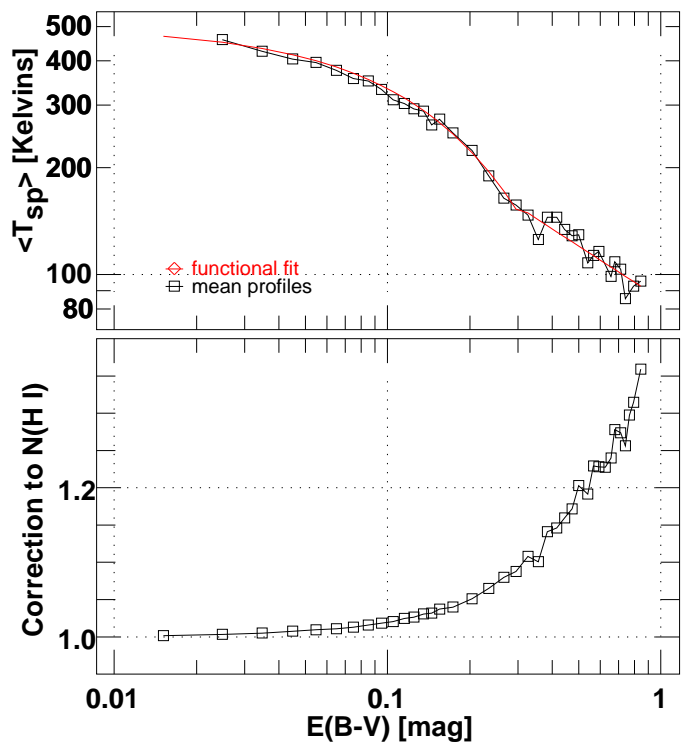}
 \caption[]{Top: mean spin temperature such that derived H I column densities 
conform to the power-law optical depth-reddening relationship shown in Fig. 3. 
The symbols represent the values for mean profiles binned in \EBV\ (see Fig. 2) 
and the smooth curve is a fitted spin temperature function that was applied in 
Fig. 1.  Bottom: The correction factor at each \EBV\ that results from applying 
the spin temperature fit to the mean profiles binned in \EBV, expressed as a 
multiplicative factor for the infinitely optically thin H I column density.
}
\end{figure}

\begin{figure*}
\includegraphics[height=8.4cm]{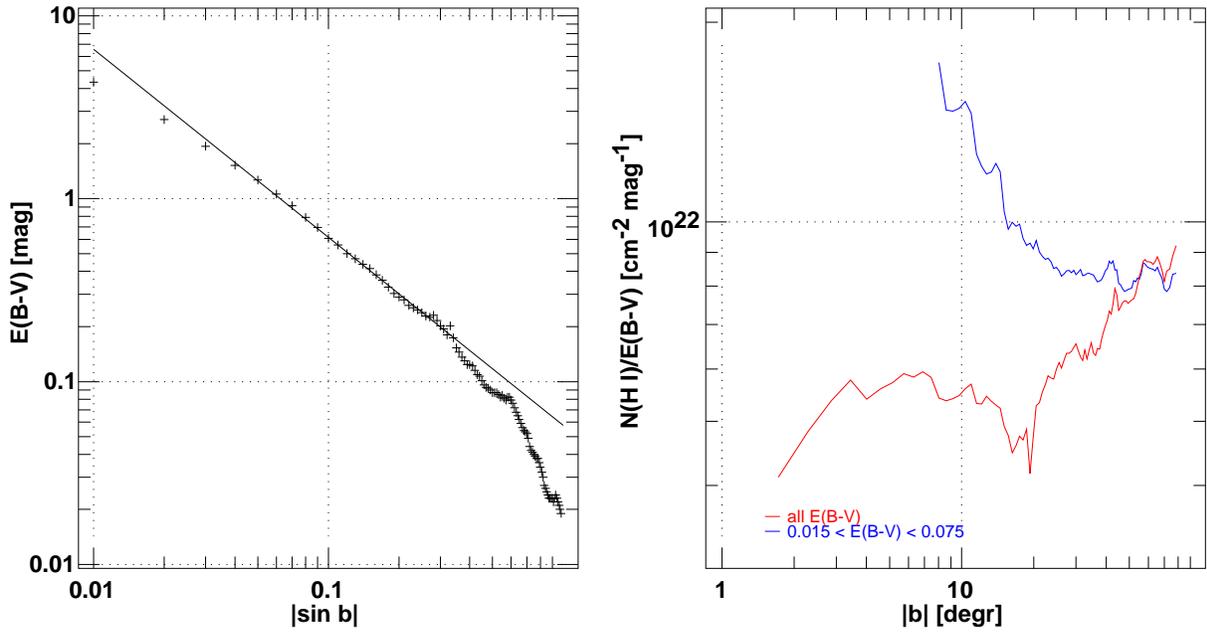}
\caption[]{Left: mean \EBV\ as a function of sin(\absb).  Points
were computed at steps of 0.01 in sin(\absb) at even 0.5\degr\ positions
around the Galaxy as in Fig. 1.  The solid line is a regression fit 
(power-law slope -1.028) to the points at 0.04 $\le$ sin(\absb) $\le$ 0.3.  
Right: N(H I)/\EBV\ averaged around the Galaxy at the latitudes used
in the panel at left. Results are shown for all \EBV\ (red, lower curve) 
and for 0.015 $\le$ \EBV\ $\le$ 0.075 mag (blue, upper curve).
}
\end{figure*}

\begin{figure}
\includegraphics[height=8.4cm]{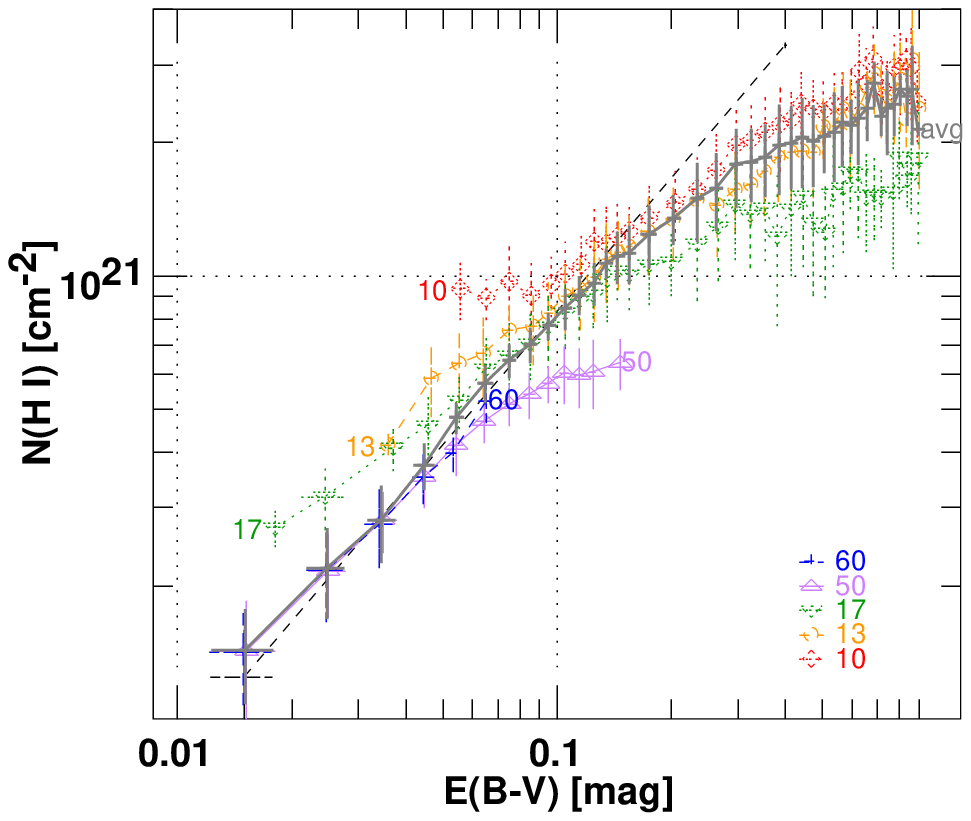}
\caption[]{Like Fig. 1 but for binned data at high and low \absb.
The mean of all data is shown as the solid black line.  The dashed  black
line corresponds to N(H I)/\EBV\ $= 8.3 \times 10^{21}\pcc$ mag$^{-1}$.
} 
\end{figure}

A general saturation correction can be determined by constraining the derived
column densities N(H I) to conform to the empirical \WHI-\EBV\ relationship, 
where the free parameter connecting N(H I) to \WHI\ is the spin temperature
that is used to convert the observed brightness temperature profiles to N(H I).
From the single power law for the \WHI-\EBV\ relationship and the inflected variation 
of N(H I) with \EBV\ in Fig. 1 it may be inferred that there is a variation in the mean 
spin temperature with increasing \EBV, actually a decline leading to a saturation 
correction that increases with \EBV\ as expected.

 We began by deriving mean \Tsp\ values from binned H I profiles (Fig. 2) across 
the range of \EBV, assuming the  power-law \WHI-\EBV\ relationship shown in Fig. 3: 
these mean \Tsp\ are shown in the upper panel of Fig. 4.  Then we fit a smooth function 
to the variation of \Tsp\ with \EBV\ and applied that to all H I profiles individually 
to correct them for the optical depth implied by their known reddening.  This process 
is self-consistent in reproducing the power-law \WHI-\EBV\ relationship when applied to 
the data at large. The power-law \WHI-\EBV\ relation breaks down at \EBV\ $<$ 
0.02 mag, N(H I) $< 2\times 10^{20}\pcc$, leading to an underestimation of
the mean \Tsp, but no correction for saturation is needed at such small N(H I) 
anyway.

Fig. 4 at bottom shows the derived saturation correction as a multiplicative 
correction  to the values of N(H I) derived from the mean H I profiles in the limit of 
zero optical depth, N(H I) $= 1.823 \times 10^{18}\pcc \int \TB dv$ 
with the integral expressed in units of K-\kms. Note that the \Tsp\ variation and the 
corrections shown are relevant only to the data that was considered.  A different dataset
might require a different variation of \Tsp\ with \EBV\ (see Section 4 below) 
and the magnitude of the correction that must be applied depends not only on \Tsp\
but on the H I profile itself. When the profile integral  is very small, even very small 
\Tsp\ do not result in a significant correction to the optically thin value of N(H I).
As well, the derived \Tsp\ at a given \EBV\ might be very different for different
datasets without implying significantly different correction factors at that \EBV.
The point is that \WHI\ $\propto$ N(H I)/\Tsp\ by definition but N(H I) depends
on \Tsp\ only when the optical depth is high.

In any case, the correction factor relevant to the dataset shown in Fig. 1 is below
20\% for \EBV\ $<$ 0.5 mag. In \cite{Lis14} we used a constant \Tsp\ = 145 K but the
important ramifications of the data are the same.  There is no appreciable correction
to N(H I) at small \EBV, leading to a reliable value for N(H I)/\EBV.  The correction
for saturation is not responsible for the inflection in the plot of N(H I)/\EBV,
which must be ascribed to the onset of \HH-formation.  

 The relevance and accuracy of using a so-called isothermal correction to N(H I) given 
the optical depth absorption profile has recently been examined by \cite{CheKan+13}, 
who derive correction factors comparable to ours.  The correction is described as
isothermal because a single \Tsp\ is applied at each velocity to emission that is
a blend of contributions from different gas phases.  The correction does an excellent 
job of bounding even very large errors in N(H I) that may occur at \WHI\ = 1-10 \kms\ 
when the optical depth is unknown. The method adopted here is broader because we use 
an implied optical depth integral \WHI\ to derive a single value of \Tsp\
across an entire line profile but the conclusions are the same.

\section{The influence of \HH\ formation on N(H I)}

The empirical correction factors derived in Sect. 2 were applied to the data in
Fig. 1 but a significant gap remains at \EBV $\ga$ 0.1 mag between the observed, 
corrected N(H I) and
the straight line N(H I)/\EBV\ $= 8.3 \times 10^{21}\pcc$ mag$^{-1}$ that is 
applicable below \EBV\ = 0.08 mag where \HH-formation does not occur 
\citep{BohSav+78}.  That is also the line of zero molecular fraction, and 
superposed on the data are (dashed) lines of constant molecular hydrogen 
fraction \fH2\ = 1-N(H I)/N(H), indicating very high molecular fractions 
at high \EBV.

Also shown in Fig. 1 are the results of a model calculation of \HH-formation in
a diffuse gas at low density n(H) = 14$\pccc$.  These are the same equilibrium
heating/cooling/\HH-formation models we have used earlier in for instance 
\cite{Lis07CO} to illustrate \HH\ and CO formation, but now with the remainder H I shown
on the vertical axis.  Also, the models were calculated using the newly inferred value 
N(H)/\EBV\ $= 8.3 \times 10^{21}\pcc$ mag$^{-1}$ that provides less dust shielding
and extinction at a given N(H).  Each leaf of the plot is for a separate cloud model
having a central hydrogen column density N(H) differing by a factor 2$^{1/4}$ from its
neighbors and the variation along each leaf represents the locus of column density
seen at all impact parameters across the face of the model.  The right-most points along 
each leaf correspond to sightlines  passing closer to the center of the model and so 
would be observed with smaller probability in real observations.

In any case, the point is to demonstrate that although there seems to be little alternative
to \HH-formation, that explanation also works in practice.

\section{Latitude variation}

Fig. 1 shows that the data at latitudes at \absb\ $\ge$ 20\degr\ fit together into 
a  coherent whole with a single message about N(H I)/\EBV\ over the Galaxy and over 
a wide range of \EBV.  This is not true of the sky at smaller \absb, as will be
discussed now.

Figure 5 at left shows the latitude variation of the mean reddening averaged around
the sky.  Below about 15\degr\ much of the sky is well-described by the cosecant law 
expected for a plane-parallel stratified medium. At higher latitudes the latitude 
dependence progressively steepens with, finally, a cubic dependence (actually, 
power-law -3.1) for \absb\ $>$
36\degr.  Given these gradients we worried that the larger H I beamsize of the LAB
survey might have artificially increased the values we derived for N(H I)/\EBV.  The 
disparity in beamsizes (36\arcmin\ vs 6\arcmin) would merely introduce scatter for a 
uniform sky, but the larger H I beam has a slightly lower intensity-weighted mean 
\absb\ when viewing a medium that is concentrated to the galactic equator.  

Numerical integration over the H I beam using the gradients shown in Fig. 5 suggested
that the effect would not be important but as a test we recalculated our results 
comparing N(H I) with \EBV\ measured 9\arcmin\ and 18\arcmin\ closer to the galactic
equator (ie 50\% and 100\% of the radius of the H I beam).   We found only that the mean 
N(H I)/\EBV\ declined progressively below \absb\ = 8\degr,  by a maximum of
5\% at \absb\ = 4\degr.  The vertical sky gradient should not have affected any of 
the conclusions drawn in this work. 

At the right in Figure 5 we show the variation in the mean N(H I)/\EBV\ at 
 0.015 $\le$ \EBV\ $\le$ 0.075 mag and over all \EBV
\footnote{ Note that we derived a separate saturation correction at \absb\ 
$<$ 20\degr\ where profiles are broader, with smaller peak brightness 
and integrated opacity at a given N(H I) or \EBV.}.  N(H I)/\EBV\ measured
over the confined range varies little at \absb\ $\ge$ 20\degr, in keeping
with Fig. 1,  but increases rapidly down to \absb\ = 8\degr,
below which there are no LAB survey grid-points in the confined \EBV\ range.  
The mean N(H I)/\EBV\ taken over all \EBV\ behaves oppositely, staying near 
N(H I)/\EBV\ = $5-6 \times 10^{21} \pcc$ mag$^{-1}$ at \absb\ $\la$ 20\degr\ 
and increasing above.  The two curves increasingly coincide at larger
\absb\ where there are fewer positions with high reddening. The values 
N(H I)/\EBV\ $= 5-6 \times 10^{21}\pcc$ mag$^{-1}$  measured over 
all \EBV\ at \absb\ $\la$ 30\degr\ may explain the difference between and
our results and those obtained previously \citep{SavDra+77,ShuVan85,DipSav94} 
from stellar absorption spectra.

To better understand the latitude variation, Fig. 6 shows 
N(H I)/\EBV\ for data at high and low \absb, binned in \EBV.  
\HH\ formation forces N(H I) down 
while the curves in Fig. 6 lay higher at smaller \absb\ where the overall \HH-fraction
should be larger.  Reconciling the low-latitude curves in Fig. 6 with a constant 
N(H)/\EBV\ $= 8.3 \times 10^{21}\pcc$ mag$^{-1}$ requires recognizing two
effects:
\begin{description}
  \item[i)]  The curves generally rise for smaller \absb\ because sightlines at lower latitude
  and higher \EBV\ are, somewhat counter-intuitively, less likely to be molecular.  Only at low latitude 
   is it possible to accumulate strong reddening over long pathlengths, in gas devoid of \HH.
   \item[ii)] The low-latitude curves lie above the regression line.  So either the 
  gas/reddening ratio increases closer to the galactic plane or the curves must be shifted 
  to the right.  Very recent results from Planck \cite{PlaXXXI13} suggest just such a correction 
as noted in Section 5.
\end{description}

\section{Discussion}

When \EBV\ from SFD98 is compared with 21cm H I column densities at  
0.015 $<$\EBV\ $<$ 0.075 mag where \HH-formation should be negligible, 
the  stable ratio N(H I)/\EBV\ = 
$8.3 \times 10^{21}\pcc$ mag$^{-1}$  derived at \absb\ $\ge$ 20\degr\ 
substantially exceeds the seemingly universal
value N(H I)/\EBV\ $= 4.8 - 5.2 \times 10^{21}\pcc$ mag$^{-1}$ that is 
derived from optical/uv absorption line work \citep{BohSav+78,MirGer79,ShuVan85,
DipSav94}.  Moreover the N(H I)/\EBV\ ratio measured over the same restricted range of
reddening increases strongly for \absb\ $\la$ 20\degr, approximately doubling
down to \absb\ = 8\degr, at which point sightlines with \EBV\ $<$ 0.075 mag
do not exist in the data we considered at the gridpoints of the LAB all-sky
H I survey,

At \absb\ $\ga$ 20\degr, corrections to N(H I) for optical depth in the 21cm line 
exceed 20\% only at \EBV\ $\ga$ 0.5 mag, given the observed profiles and the
integrated H I optical depth-reddening law we derived from interferometric
H I absorption measurements,  $\int \tau(H I) dv = 14.07 ~\kms\ $\EBV$^{1.074}$ 
at 0.02 $\la$ \EBV\ $\la$ 3 mag. Optical depth corrections to N(H I) are
even smaller at  10\degr\ $\la$ \absb\ $\la$ 20\degr\ where the galactic velocity
gradient is more apparent and profiles have smaller peak 21cm brightness
for a given N(H I) and \EBV.  In this case only \HH-formation can be
responsible for the inflection in the N(H I)-\EBV\ relationship that occurs
at \EBV\ $\ga$ 0.08 mag and we showed model results for \HH-formation in
a low density diffuse gas that are a good match to the data.

In our earlier work we discussed possible corrections to the reddening
maps of SFD98, based on previous work by other investigators; the 
prevailing view appeared to be that the reddening maps of SFD98,
had, if anything, overestimated \EBV.  Correcting the results of
SFD98 in  that manner  would only exaggerate the effect discussed here.  
More recently the Planck dust maps have appeared \citep{PlaXXXI13}, 
and they may tell
a different story.  Although the Zodaical Emission-corrected 353 GHz
optical depth maps converted to reddening give the same result we
derived, N(H I)/\EBV\ $= 8.3 \times 10^{21}\pcc$ mag$^{-1}$, those
maps are recommended for use only at larger \EBV. By contrast,
the Planck reddening maps based on QSO colors that are recommended for use
below \EBV\ = 0.3 mag would have \EBV$^\prime$ = (\EBV$_{SFD}$+0.003 mag)/0.92.
When such a transformation is used to rederive the present results,
we find a high-latitude asymptote 
N(H I)/\EBV\ $= 7.2 \times 10^{21}\pcc$ mag$^{-1}$, accounting
for about half the effect noted, in the log sense. This 
transformation increasing the \EBV\ values of SFD98 at smaller
\EBV\ would shift the curves in Fig. 6 to the right in the manner
required to reconcile them with a single value of N(H I)/\EBV, as  
discussed in Sect. 4.


\acknowledgments

  The National Radio Astronomy Observatory is a facility of the National
  Science Foundation operated under contract by Associated  Universities, Inc.
  The author was partially funded by the grant ANR-09-BLAN-0231-01 from the 
  French {\it Agence Nationale de la Recherche} as part of the SCHISM 
  project (http://schism.ens.fr/). The author thanks Maryvonne Gerin
  for her hospitality at the ENS where this manuscript was finished.
  The author also thanks an anonymous referee for comments that served
  to clarify and improve the discussion.




\bibliographystyle{apj}


\begin{thebibliography}{16}
\expandafter\ifx\csname natexlab\endcsname\relax\def\natexlab#1{#1}\fi

\bibitem[{{Bohlin} {et~al.}(1978){Bohlin}, {Savage}, \& {Drake}}]{BohSav+78}
{Bohlin}, R.~C., {Savage}, B.~D., \& {Drake}, J.~F. 1978, ApJ, 224, 132

\bibitem[{{Chengalur} {et~al.}(2013){Chengalur}, {Kanekar}, \&
  {Roy}}]{CheKan+13}
{Chengalur}, J.~N., {Kanekar}, N., \& {Roy}, N. 2013, MNRAS, 432, 3074

\bibitem[{{Dickey} {et~al.}(1983){Dickey}, {Kulkarni}, {Heiles}, \& {Van
  Gorkom}}]{DicKul+83}
{Dickey}, J.~M., {Kulkarni}, S.~R., {Heiles}, C.~E., \& {Van Gorkom}, J.~H.
  1983, ApJS, 53, 591

\bibitem[{{Diplas} \& {Savage}(1994)}]{DipSav94}
{Diplas}, A. \& {Savage}, B.~D. 1994, ApJ, 427, 274


\bibitem[{{Kalberla} {et~al.}(2005){Kalberla}, {Burton}, {Hartmann}, {Arnal},
  {Bajaja}, {Morras}, \& {P{\"o}ppel}}]{KalBur+05}
{Kalberla}, P.~M.~W., {Burton}, W.~B., {Hartmann}, D., et al. 2005, A\&A, 440, 775

\bibitem[{{Kanekar} {et~al.}(2011){Kanekar}, {Braun}, \& {Roy}}]{KanBra+11}
{Kanekar}, N., {Braun}, R., \& {Roy}, N. 2011, ApJ, 737, L33

\bibitem[{{Liszt}(2014)}]{Lis14}
{Liszt}, H. 2014, ApJ, 780, 10

\bibitem[{{Liszt}(2007)}]{Lis07CO}
{Liszt}, H.~S. 2007, A\&A, 476, 291

\bibitem[{{Liszt} {et~al.}(2010){Liszt}, {Pety}, \& {Lucas}}]{LisPet+10}
{Liszt}, H.~S., {Pety}, J., \& {Lucas}, R. 2010, A\&A, 518, A45

\bibitem[{{Mirabel} \& {Gergely}(1979)}]{MirGer79}
{Mirabel}, I.~F. \& {Gergely}, T.~E. 1979, A\&A, 77, 110

\bibitem[{{Peek} {et~al.}(2011){Peek}, {Heiles}, {Douglas}, {Lee}, {Grcevich},
  {Stanimirovi{\'c}}, {Putman}, {Korpela}, {Gibson}, {Begum}, {Saul},
  {Robishaw}, \& {Kr{\v c}o}}]{PeeHei+11}
{Peek}, J.~E.~G., {Heiles}, C., {Douglas}, K.~A., et al. 2011, ApJS, 194, 20

\bibitem[{{Planck Collaboration} {et~al.}(2013){Planck Collaboration},
  {Abergel}, {Ade}, {Aghanim}, {Alina}, {Alves}, {Armitage-Caplan}, {Arnaud},
  {Ashdown}, {Atrio-Barandela}, \& et~al.}]{PlaXXXI13}
{Planck Collaboration}, {Abergel}, A., {Ade}, P.~A.~R., et al.2013, arXiv preprint, 
 astroph:1312.1300, submitted to A\&A

\bibitem[{{Roy} {et~al.}(2013){Roy}, {Kanekar}, {Braun}, \&
  {Chengalur}}]{RoyKan+13}
{Roy}, N., {Kanekar}, N., {Braun}, R., \& {Chengalur}, J.~N. 2013, MNRAS,436, 2352

\bibitem[{{Savage} {et~al.}(1977){Savage}, {Drake}, {Budich}, \&
  {Bohlin}}]{SavDra+77}
{Savage}, B.~D., {Drake}, J.~F., {Budich}, W., \& {Bohlin}, R.~C. 1977, ApJ,
  216, 291

\bibitem[{{Schlegel} {et~al.}(1998){Schlegel}, {Finkbeiner}, \&
  {Davis}}]{SchFin+98}
{Schlegel}, D.~J., {Finkbeiner}, D.~P., \& {Davis}, M. 1998, ApJ, 500, 525

\bibitem[{{Shull} \& {van Steenberg}(1985)}]{ShuVan85}
{Shull}, J.~M. \& {van Steenberg}, M.~E. 1985, ApJ, 294, 599

\end{thebibliography}


\end{document}